\documentclass[aps,twocolumn,showpacs]{revtex4}
\usepackage{graphicx,bm}
\usepackage{amsmath}
\usepackage{mathrsfs}
\usepackage{color}
\usepackage[utf8]{inputenc}
\usepackage{hyperref}
\usepackage{amsfonts}

\usepackage{color}
\definecolor{shiningblue}{rgb}{0.3,0.68,0.89}
\definecolor{pastel_red}{rgb}{1,0.41,0.38}
\definecolor{pastel_green}{rgb}{0.46,0.82,0.46}
\definecolor{pastel_blue}{rgb}{0.47,0.7,0.9}
\definecolor{soft_blue}{rgb}{0,0,0}

\def \>{\rangle} 
\def \<{\langle} 
 
\def\be{\begin{equation}} 
\def\ee{\end{equation}}

\newcommand \bea {\begin{eqnarray}} 
\newcommand \eea {\end{eqnarray}}

\begin{document}

\title{A thermodynamic paradigm for solution demixing inspired by nuclear transport in living cells}

\author{Ching-Hao Wang}
\email{chinghao@bu.edu}
\author{Pankaj Mehta}
\email{pankajm@bu.edu}
\affiliation{Department of Physics, Boston University, Boston, MA 02215}
\author{Michael Elbaum}
\email{michael.elbaum@weizmann.ac.il}
\affiliation{Department of Materials and Interfaces, Weizmann Institute of Science, Rehovot, Israel}

\date{\today}


\begin{abstract}

Living cells display a remarkable capacity to compartmentalize their functional biochemistry. A particularly fascinating example is the cell nucleus. Exchange of macromolecules between the nucleus and the surrounding cytoplasm does not involve traversing a lipid bilayer membrane. Instead, large protein channels known as nuclear pores cross the nuclear envelope and regulate the passage of other proteins and RNA molecules. Beyond simply gating diffusion, the system of nuclear pores and associated transport receptors is able to generate substantial concentration gradients, at the energetic expense of guanosine triphosphate (GTP) hydrolysis. In contrast to conventional approaches to demixing such as reverse osmosis or dialysis, the biological system operates continuously, without application of cyclic changes in pressure or solvent exchange. Abstracting the biological paradigm, we examine this transport system as a thermodynamic machine of solution demixing. Building on the construct of free energy transduction and biochemical kinetics, we find conditions for stable operation and optimization of the concentration gradients as a function of dissipation in the form of entropy production. 
\end{abstract}
\maketitle


\begin{figure}
\center{
\includegraphics[width=9cm]{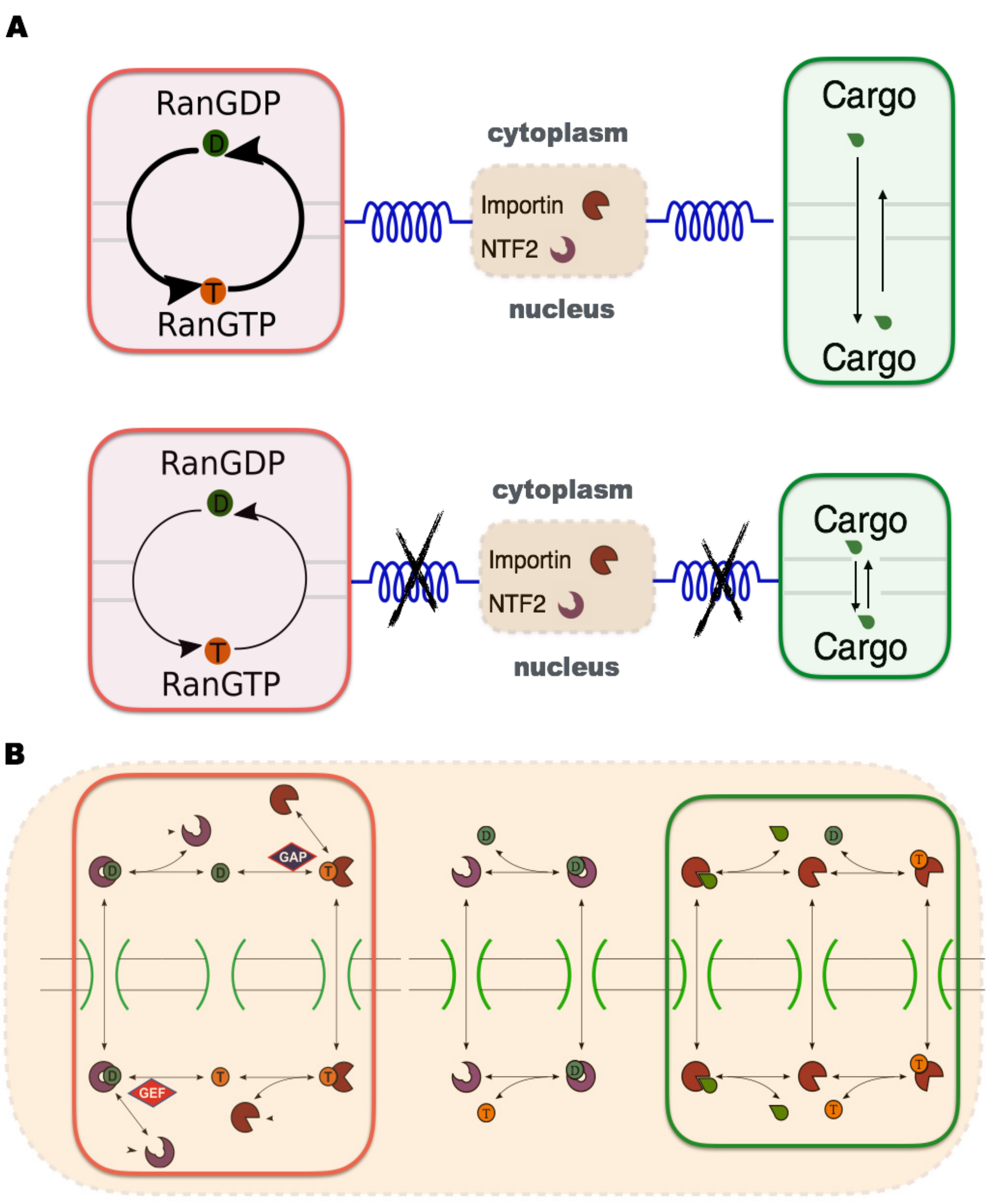}}
\caption{(Color online)  Demixing of cargo across the nuclear membrane is driven by Ran coupled to NTF2 and importin system. (A) With such coupling (upper panel), nuclear cargo accumulation is favored and Ran GTP/GDP exchange cycle proceeds faster than without coupling (lower panel). The thickness of arrowed curves in Ran cycle indicates strength of reaction flux; length of arrowed lines in cargo transport represents the rate at which the underlying processes occur. (B) Details of molecular demixing machine in the context of nuclear transport. Species labels as above. Reactions corresponding to Ran cycle and cargo transport are highlighted by red and green boxes, respectively. The orange dashed box includes all reactions coupled by the importin-NTF2 system. See also Fig S1 in the Supplementary Material.}\label{fig1}
\end{figure}


Demixing of solutions is a difficult thermodynamic problem with important practical consequences \cite{dijkstra1994evidence}. Examples include the desalination of seawater, medical dialysis, and chemical purification. In all of these processes, free energy is consumed in order to balance entropy of mixing. Typical engineering approaches to demixing involve application of hydrostatic pressure (reverse osmosis), solution exchange (dialysis), or phase change (crystallization or distillation) \cite{mistry2011entropy,glynn1990phasequi}. In this context living cells adopt a fundamentally different paradigm by establishing and maintaining concentration gradients at \emph{steady-state} under a fixed set of intrinsic thermodynamic parameters. This recalls the similar capacity to operate mechanochemical motors isothermally \cite{isomotor:Andrea,isomotor:Parrondo}.

A prominent example of molecular separation is the eukaryotic cell nucleus, wherein the concentrations of many proteins and RNA differ significantly from those in the cell body (cytoplasm). These gradients are maintained by a transport system that shuttles molecular cargo in and out via large protein channels known as nuclear pores \cite{maul1977qNP,talcott1999crossNE}. This system has been under intensive study in the biological \cite{gorlich1999karyopherin,stewart2007molecular:NatureReview,Hetzer2008CellReview_NPC,wente2010ColdSpringReview,kimura_biological_2014} and biophysical  \cite{peters_optical_2003,michaelPNAS,wagner_promiscuous_2015,zahn_physical_2016,vovk_simple_2016} literatures, with particular emphasis on single-molecule interactions at the pore itself \cite{keminer_permeability_1999,yang2004imaging,kubitscheck2005nuclear,grunwald_nuclear_2011}. Simple thermodynamic considerations make clear that equilibrium pore-molecule interactions are insufficient to support concentration gradients in solution. Demixing between two compartments cannot occur spontaneously, but must be coupled to a free energy source \cite{hill2012}. At the same time, demixing does not require rectified translocation \cite{gorlich2003characterization}. Concentration gradients may be established in the presence of a balanced, bi-directional exchange \cite{michaelPNAS,michaelHSPJ,lolodi_dissecting_2016}.

Nuclear pores represent an unusual transporter in that there is no membrane to cross. Water, ions, and small molecules diffuse freely across the nuclear envelope to equilibrate between the two compartments. Generally, the permeability drops between molecular weight 20 kDa and 40 kDa \cite{peters1984nucleo,samudram_passive_2016}. Transport of larger macromolecules relies on a special class of proteins, called transport receptors  (i.e. ``importin''), that usher their cargoes across the nuclear pores by virtue of specific interactions with the channel components. Recognition between importins and their molecular cargo depends on the presence of particular amino acid sequences known as nuclear localization signals (NLS) \cite{grote1995mapping,gorlich1999karyopherin,rexach1995protein}. The affinity between importin and cargo is regulated by a small GTP-binding protein called Ran \cite{SmithScience,terry2007crossing:science}. When associated with GTP (RanGTP), Ran binds strongly to importin in a manner that is competitive to NLS binding. By contrast, Ran associated with GDP (RanGDP) binds importin very weakly. Ran interconverts between these two forms through GTP hydrolysis and GTP/GDP exchange, facilitated by the GTPase Activating Protein (RanGAP) and the Guanosine Exchange Factor (RanGEF), respectively  \cite{bos2007gefsfreeratio}. RanGAP is structurally bound to the cytoplasmic face of the nuclear pore  and RanGEF is bound to chromatin. Their activities generate a high concentration of RanGTP in the nucleus and RanGDP in the cytoplasm (see FIG.\ref{fig1}). 

Demixing is powered by transducing free energy from GTP hydrolysis through the interactions of transport receptor with Ran. The transport machinery has been formulated in terms of coupled chemical kinetics \cite{gorlich2003characterization,riddick_systems_2005,michaelBiophyJ} but the energetics have not yet been addressed. In particular, we ask: How does the rate of dissipation (energy consumption) relate to the achieved concentration gradient? What is the proper definition of transport efficiency? Is there an optimal working point given the nonequilibrium nature of this cellular machine? To address these questions, it is helpful to reformulate the problem in a thermodynamic language. For consistency with the literature we retain the biological nomenclature, yet the aim is to understand the natural engineering in a more abstract sense that might ultimately be implemented synthetically.

In the thermodynamic formulation, a central role is played by energy transduction in a ``futile cycle'' among the components (see FIG.~\ref{fig1}). This is roughly analogous to heat flow in a Carnot cycle. The importin receptor binds RanGTP, and a second receptor known as nuclear transport factor 2 (NTF2) binds specifically RanGDP. The forward cycle takes RanGTP out to the cytoplasm with importin and RanGDP back to the nucleus with NTF2. Detailed balance is broken by the distribution of RanGAP and RanGEF as described above, so that the reverse cycle is scarcely populated. 

Free energy from the Ran cycle is transduced by importin to bias the steady-state free cargo concentrations in the nuclear and cytoplasmic compartments. Details of the underlying biochemical reactions are shown in FIG.~\ref{fig1}B and can be modeled on the basis of mass action. The corresponding kinetic parameters can be found from the literature or estimated from simple scaling arguments (see FIG.~S1 and Supplementary Material for details of the kinetic model\cite{supplementary_ref}). Numerical solutions are obtained by solving all the coupled rate equations using a standard Runge-Kutta method (The code used for simulation is available in the Supplementary Material). We emphasize that the present aim is not so much to model the biological implementation as to explore the generic operation of the thermodynamic machine. Relations between parameters are therefore more important than specific values.

Energetics enter the model via the charging of Ran with GTP and its subsequent hydrolysis to GDP (reactions 5  and 2 in FIG.~S1, respectively). The flux through these two reactions must be equal in steady state. Energy is drawn from the non-equilibrium ratio of free GTP to GDP, $\theta$, which is maintained by cellular metabolism and defines an effective ``free energy'' $F_\theta :=k_B T\log \left(\theta\right)$. A typical value of $\theta$ is roughly a few tens to a hundred  \cite{biochemtext, bos2007gefsfreeratio}. Independent of the complex operational details of RanGEF and RanGAP with associated co-factors, we can look at the steady states and relate the reactions to $\theta$. (See Supplementary Material for details.) On the nuclear side, the complex NTF2-RanGDP exchanges for NTF2 and RanGTP. The dissociation constant $K_D$ (forward divided by reverse flux) can be shown to be proportional to $\theta$. Conversely, on the cytoplasmic side the corresponding $K_D$ is proportional to $1/\theta$. As a result, any enhancement of flux through the futile cycle in the forward reaction conferred by increasing $\theta$ (i.e. reaction 5 in FIG.S1C) is balanced by the contradicting counterpart in preventing RanGTP release (i.e. reaction 2 in FIG.S1C). 

A useful measure of cargo demixing is the nuclear localization ratio, NL, defined as the ratio between nuclear and cytoplasmic cargo concentrations: $[C]_{nu}/[C]_{cyto}$. This ratio defines a chemical potential, $\Delta\mu = -k_BT \log{[C]_{nu}/[C]_{cyto}}$, that measures the excursion from equilibrium. FIG. \ref{fig2}A shows NL as a function of importin and NTF2 concentrations. The most striking feature is that NL is maximum for intermediate levels of importin. The importin concentration at which NL is maximized, $[Im^*]$, grows with the total cargo load, $[C]_{tot}$ (see FIG. \ref{fig2}B). Furthermore, $[Im^*]$ is largely independent of NTF2 concentration for different cargo concentration considered (see FIG. S5). This suggests an inherent optimization.


\begin{figure}
\begin{center}
\includegraphics[width=9cm]{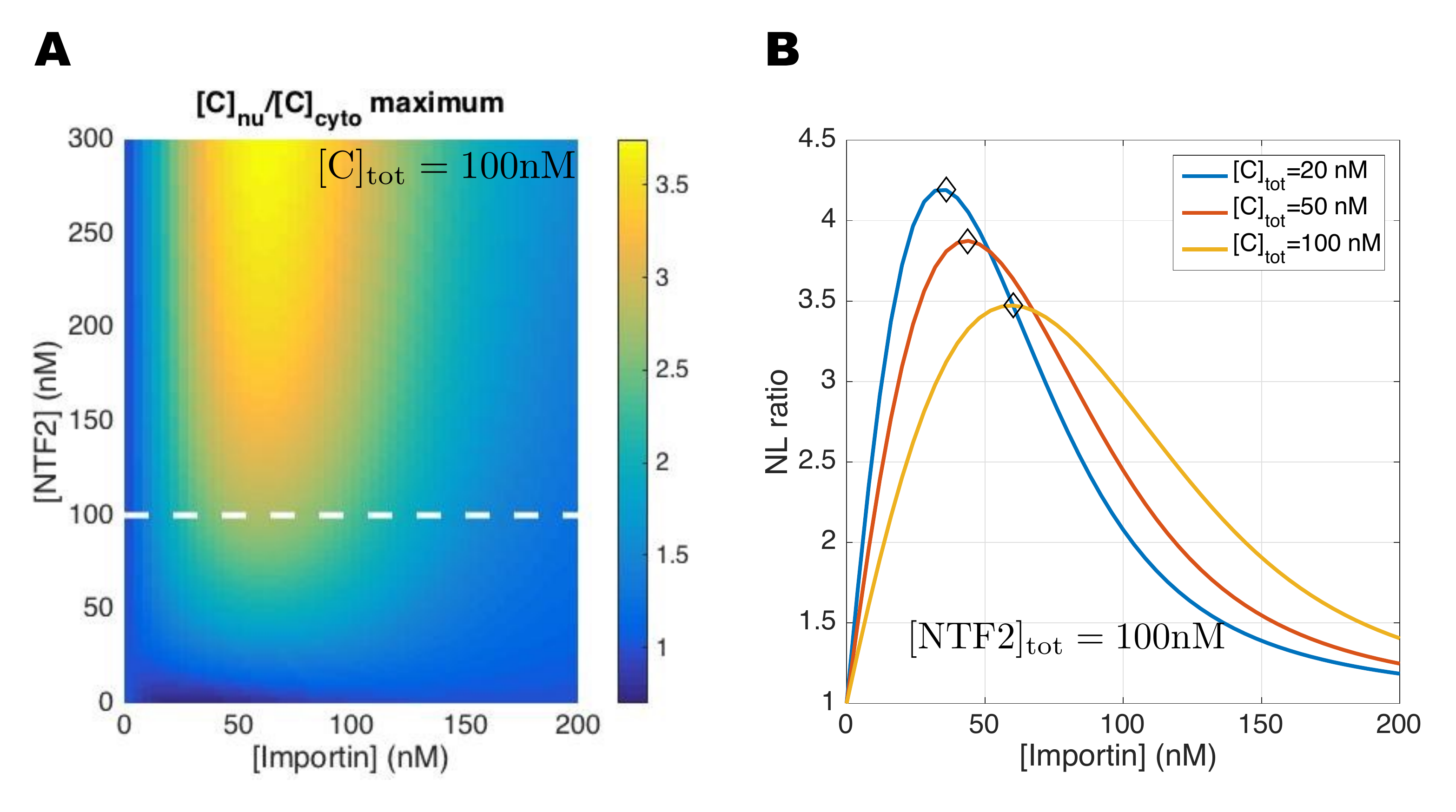}
\caption{(Color online)  Phase diagram of nuclear localization. (A) The cargo nuclear localization NL$:=[C]_{nu}/[C]_{cyto}$ (color shadings) is obtained by varying the total importin and total NTF2 concentrations while keeping overall cargo level fixed at $[C]_{tot}=100$ nM. (B)  A family of curves shows NL for several cargo concentrations as a function of importin concentration with $[NTF2]_{tot}=100$ nM. The 1D curve for $[C]_{tot}=100$ nM is a cut across the plot of panel A. Locations of NL maximum are marked by diamonds (see FIG.\ref{fig4}C as well). Kinetic rate constants used are given in the Supplementary Material. Total Ran concentration $[Ran]_{tot}$=75 nM. }\label{fig2}
\end{center}
\end{figure}

At first sight it is surprising that augmenting the importin concentration, which increases the number of molecules that can transport cargo to nucleus, may decrease the localization ratio. The optimal dependence of NL on importin reflects the dual role importin plays as the \emph{inbound carrier of cargo protein} as well as the \emph{outbound carrier of RanGTP}. Powering the futile cycle requires that importin bind RanGTP, whereas cargo transport requires importin to bind cargo. This establishes a binding competition in the nucleus that is a characteristic feature of protein import (FIG.\ref{fig3}A). In spite of the higher affinity of RanGTP for importin, the cycle analysis shows that importin in the nucleus binds cargo more rapidly. As seen in FIG.~\ref{fig3}BC, NL is maximized close to the point at which the difference between the reaction fluxes of importin-cargo formation ( $\Phi_7^-:=\tilde{\Phi}_7^- [{Im}]_{nu}=(k_7^-[C]_{nu})[{Im}]_{nu}$ ) and importin-RanGTP formation ($\Phi_4^+:=\tilde{\Phi}_4^+[{Im}]_{nu}=(k_4^+[RanGTP]_{nu})[{Im}]_{nu}$ ) is maximal. Intuitively, this is the realm where importin can bind cargo effectively while maintaining its coupling to the reaction cycle that transduces energy for cargo transport.


\begin{figure}
\center{
\includegraphics[width= 9cm]{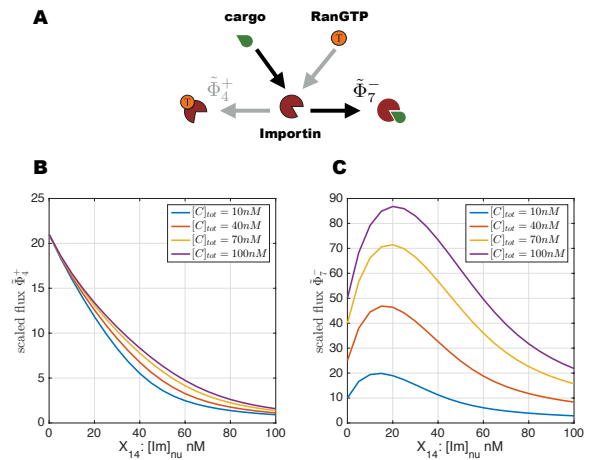}}
\caption{(Color online)  Competition between RanGTP and cargo to bind importin. (A) Schematic of the two competing reactions.  (B) Reaction flux for importin-RanGTP formation $\tilde{\Phi}_4^+\sim k_4^+[RanGTP]$ and (C) flux for importin-cargo formation $\tilde{\Phi}_7^-\sim k_7^-[C]_{nu}$. Fluxes are scaled by $[{Im}]_{nu}$ (see text). Parameters as in FIG.\ref{fig2} }\label{fig3}
\end{figure}


To understand the thermodynamics of nuclear transport, we formulate the transport system as a nonequilibrium Markov process. Since a nonequilibrium steady state (NESS) necessarily breaks detailed balance in the underlying Markov process, the system has a nonzero entropy production  \cite{hill2012,mehta2012energetic,lang2014thermodynamics}. This is the energy per unit time required to maintain the NESS, with units of power. Following the Schnakenberg description, the EP for a NESS is given by \cite{NESSref}
\be\label{EP}
EP=k_B T\sum_{i,j}P_i^{SS}W(i,j)\log\frac{W(i,j)}{W(j,i)},
\ee 
where $P_i^{SS}$ is the steady state probability distribution of state $i$ while $W(i,j)$ denotes the transition probability from state $i$ to state $j$. Concretely, $P_i^{SS}$ is the fraction of reactants that participate in the transition reaction starting from state $i$ while $W(i,j)$ can be calculated from the relevant reaction fluxes. Note that the sum in Eq.\eqref{EP} is taken over all links of the reaction network. This is equivalent to summing over the links pertaining to the Ran futile cycle. (See Supporting Material for details).

This entropy production provides a direct measure of the power input to the underlying biochemical circuit. FIG.~\ref{fig4}A shows EP for various importin and NTF2 concentrations. FIG.~\ref{fig4}B adds various cargo concentrations for a fixed level of $[NTF2]$. In each case, as the importin concentration increases, EP first drops to a minimum and then peaks before slowly decaying. Note that the minimal dissipation (entropy production) tracks closely with the value at which the NL ratio peaks (see FIG.\ref{fig4}C). These conditions define an optimal efficiency of the demixing machine. With further increasing importin concentration, the futile cycle decouples from cargo translocation and EP increases. At a still higher concentration, EP peaks and then decreases. This can be understood qualitatively as a short circuit via reaction 8, where importin moves between compartments carrying neither cargo nor RanGTP. As seen in FIG.\ref{fig4}D, for such high importin levels the corresponding flux $\Phi_8$ exceeds that of the RanGTP loading to importin, $\Phi_4$.


\begin{figure}
\begin{center}
\includegraphics[width=8.5cm]{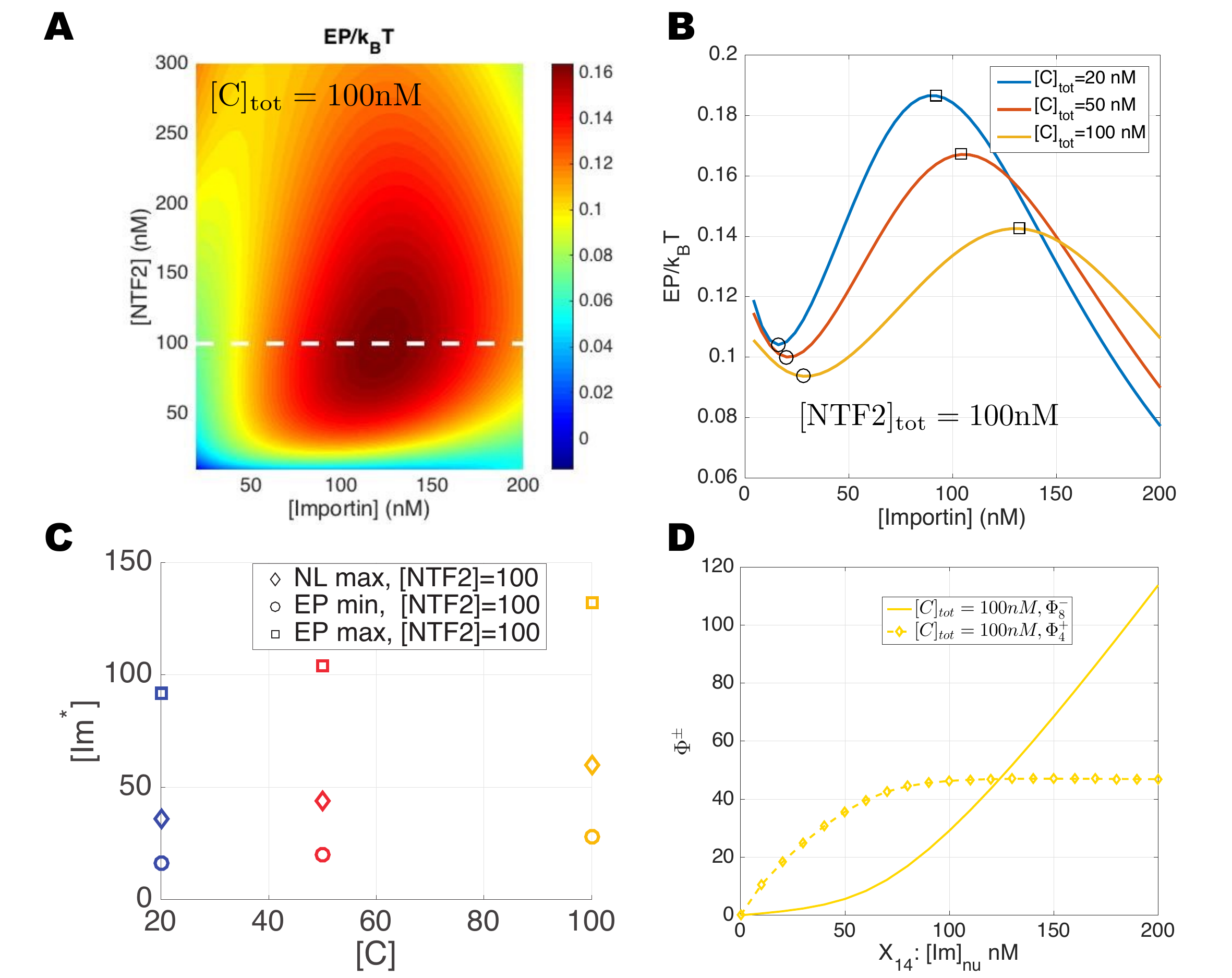}
\caption{(Color online) Phase diagram of entropy production. (A) Entropy production is plotted as a 2D function of NTF2 and importin, at fixed cargo concentration $[C]_{tot}=100$ nM. Compare with FIG. 2A. (Axes extend to 5 instead of 0 nM to avoid numerical divergence.) (B) A family of curves shows the entropy production for several cargo concentrations as a function of importin concentration; NTF2 concentration fixed at $[NTF2]_{tot}=100$ nM. The 1D curve for $[C]_{tot}=100$ nM is a cut across the plot of panel A. Compare with FIG. 2B. Peaks and troughs are marked by squares and circles, respectively. (C) Locations of entropy production maximum/minimum (square/circle) and that of nuclear localization maximum (diamond). Colors match curves in panel B. The importin concentration at which EP is minimum is close to but always less than $[Im^*]$, where NL is maximum. Thus, one strategy for maximizing the efficiency of demixing is to have the futile cycle operate in regime where its entropy production is minimized. (D) EP decreases at very high importin concentration. This reflects a loop around the energetic reaction $\Phi_4$ via the reversible reaction $\Phi_8$. Here $[NTF2]_{tot}$=100 nM, as in the panel B. In all panels, $[Ran]_{tot}=75$ nM. Kinetics constants as in FIG.\ref{fig2} and \ref{fig3} (see SM Section II).}\label{fig4}
\end{center}
\end{figure}


To the best of our knowledge the optimal steady-state has not been observed experimentally. The \emph{kinetic} rate of nuclear protein uptake was found to be reduced by microinjection of importin receptor to live cells; rate equation simulations done in parallel also pointed to the dual role of importin (FIG.~\ref{fig3}A) \cite{Ribbeck2001kinetics}. Steady-states were not reported in that study, however. Other possible experimental tests include titration of importin protein to nuclei reconstituted in vitro in \emph{Xenopus} egg extract and optical activation of importin receptors, similarly to induction of nuclear transport by NLS activation \cite{niopek_engineering_NatComm_2014}. An important point in comparison with literature is that we have considered a single, collective ``cargo" for transport. In reality, multiple cargoes compete for binding to relatively few but promiscuous transport receptors. This competition leads to a partitioning according to equilibrium binding affinities and may lead to vastly different kinetics. However the steady-state NL ratio (in solution) is independent of the affinity, reflecting thermodynamic control and equilibration of the chemical potentials \cite{michaelHSPJ,michaelBiophyJ,lolodi_dissecting_2016}. Consistent with this paradigm, in which a net accumulation occurs together with a balanced bidirectional flux, the simulations show that the nuclear and cytoplasmic concentrations of the importin-cargo complex ($X_4$ and $X_{11}$, respectively) equilibrate in steady-state. It is also interesting to note that RanGTP loading onto importin (reaction 4) was identified in the earlier analysis as the primary rate-limiting step in accumulation kinetics \cite{michaelBiophyJ}.

In summary, we have analyzed the biological paradigm for nuclear transport from a thermodynamic point of view. Building upon prior understanding that protein cargo demixing is facilitated by hydrolysis of GTP, we draw the connection between consumption of chemical energy and maintenance of the cargo concentration gradient at non-equilibrium steady states. We show that the efficacy of nuclear localization ratio peaks at intermediate importin level, which is not far from the power consumption (entropy production) minimal. It is likely that the cell maintains an importin concentration at an advantageous level with respect to these operating points defined by the thermodynamic analysis. Interestingly, the system as configured is robust to the quality of the chemical energy source, in the sense that the NL ratio is almost independent of the GTP:GDP ratio $\theta$ when $\theta \gtrsim 20$, FIG.~S4. A thermodynamic definition of the system efficiency remains elusive, however. Whereas conventional efficiency of an engine is a dimensionless ratio of mechanical to thermal power, in the NESS a constant free energetic gradient (chemical potential in the present case) is maintained by a constant power input. The ratio has units of time. This could be renormalized sensibly by a characteristic remixing time, e.g., the permeability of the nuclear pores to the cargo-importin complex. There is no guarantee of a bound at unity, however, so the definition remains ad hoc, a useful figure of merit. It is also interesting to contrast the competitive interactions between receptor and RanGTP in nuclear protein accumulation (import) with the cooperative interactions in nuclear protein depletion (export). While these are often considered as simple inverse processes, they differ in this essential aspect \cite{kim_enzymatically_2013}.

This work is part of a larger literature that seeks to examine basic biophysical processes from a thermodynamic perspective. It is now clear that thermodynamics fundamentally constrains the ability of cells to perform various task ranging from detecting external signals \cite{berg1977physics,endres2009maximum,mora2010limits}, to adaptation \cite{sartori2014thermodynamicAdaptation}, to making fidelity decisions \cite{lang2014thermodynamics}, generating oscillatory behavior \cite{elowitz2000oscillatory}, and of course generating forces and dynamic structures \cite{ndlec1997self,surrey2001organization,karsenti2008selfReview}. In all these examples, it is possible to map these tasks to Markov processes and compute the corresponding entropy production rate. This suggests that there may be general theorems about thermal efficiency in cells that are independent of the particular task under consideration  \cite{mehta2012energetic,mehta2016landauer,barato2015thermodynamic,gingrich2016dissipation}.  It will be interesting to explore if this is actually the case and to see if these principles can be applied to synthetic biology and ultimately biomimetic engineering \cite{mehta2016landauer}.

\emph{Acknowledgment}
PM and CHW were supported by a Simons Investigator in the Mathematical Modeling of Living Systems grant, a Sloan Fellowship, and NIH Grant No. 1R35GM119461 (all to PM). ME acknowledges a grant from the Israel Science Foundation (1369/10), the Gerhardt M.J. Schmidt Minerva Research Foundation, and the historical generosity of the Harold Perlman family. Simulations were carried out on the Shared Computing Cluster (SCC) at BU.

\bibliography{references.bib}{}
\bibliographystyle{apsrev4-1}

\widetext
\newpage
\begin{center}
\textbf{\large Supplemental Material for ``A thermodynamic paradigm for solution demixing inspired by nuclear transport in living cells"}
\end{center}

\author{Ching-Hao Wang}
\email{chinghao@bu.edu}
\author{Pankaj Mehta}
\email{pankajm@bu.edu}
\affiliation{Department of Physics, Boston University, Boston, MA 02215}
\author{Michael Elbaum}
\email{michael.elbaum@weizmann.ac.il}
\affiliation{Department of Materials and Interfaces, Weizmann Institute of Science, Rehovot, Israel}

\date{\today}

\maketitle
\setcounter{equation}{0}
\setcounter{figure}{0}
\setcounter{table}{0}
\setcounter{page}{1}
\makeatletter
\renewcommand{\theequation}{S\arabic{equation}}
\renewcommand{\thefigure}{S\arabic{figure}}
%
\section{Biophysical models}

\begin{figure*}[!h]
\begin{center}
\includegraphics[width=16cm]{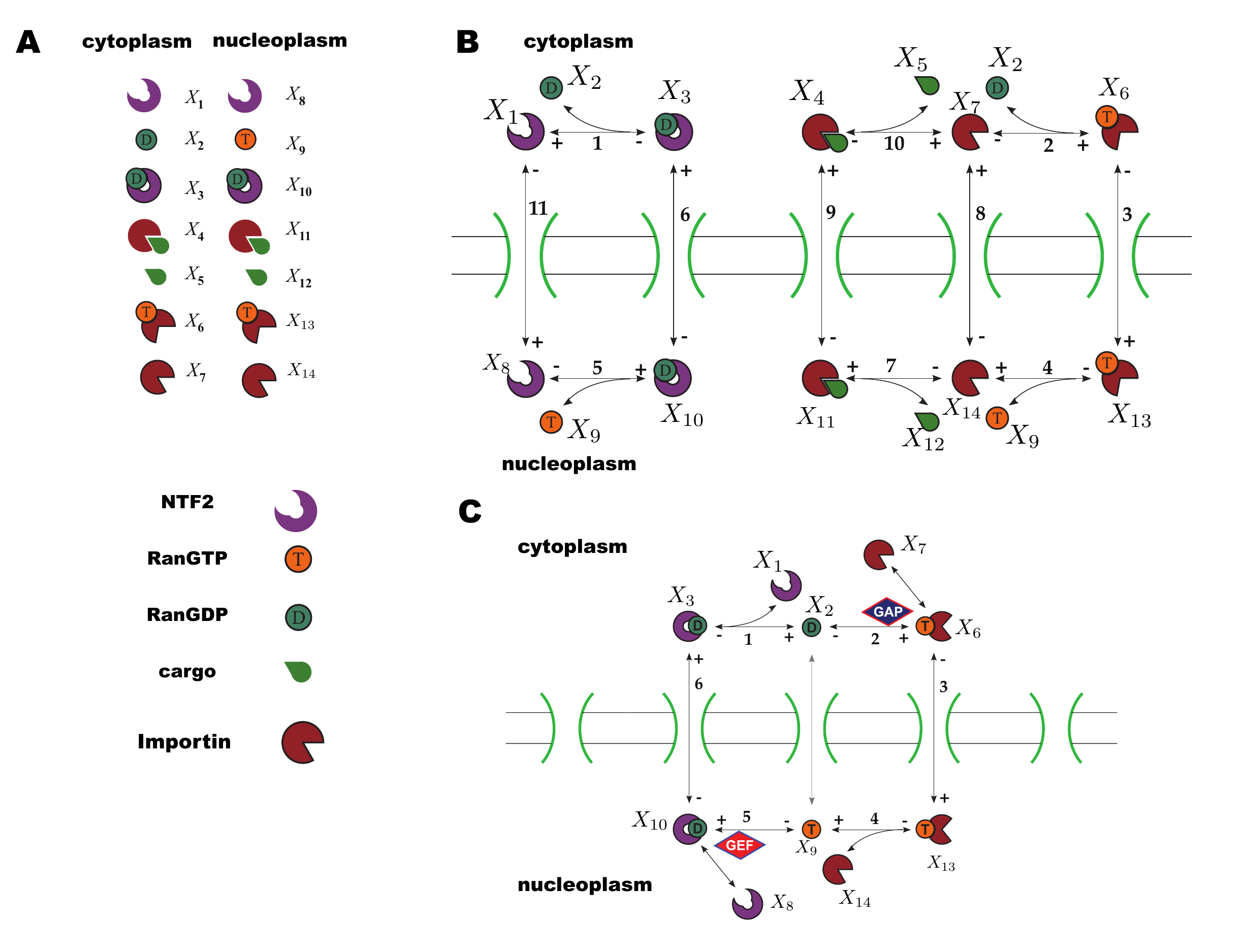}
\caption{{\bf Molecular reactions involved in nuclear transport}. (A) Molecular species in our nuclear transport model. (B) Schematics of the reactions involved in nuclear transport. Note that + and - signs represent self-consistently the start and end points of the reactions, rather than forward or reverse cycle orientations. Thus for example $k_{"+"}$ and $k_{"-"}$ for reaction 9 in Eq.(S15) take signs for loss and gain, respectively. (C) The subset of reactions in (B) that forms the futile cycle (i.e. the energy source for nuclear transport).}
\label{fig:model}
\end{center}
\end{figure*}

The 11 basic reactions constituting the whole transport process are depicted in Figure~\ref{fig:model}. Our model incorporates the known (simplified) mechanism of nuclear transport of cargo through binding with importin and the active consumption of energy through hydrolysis of GTP. Such process is facilitated by Ran's intrinsic GTPase activity, which is activated via interaction with the Ran GTPase activating protein (RanGAP). In addition, we also include the reverse conversion of RanGDP to RanGTP through the action of guanine Exchange Factor RCC1 (known as RanGEF). In addition to the \emph{standard model} of nuclear transport whose biochemistry is summarized below, we also incorporates the backward reactions to account for the reversibile nature of this transport process \cite{michaelBiophyJ}.

\begin{itemize}
\item (Reaction 10, 9) In the cytoplasm, say, compartment $A$, the complex formed by importin protein (transport receptor) and the cargo $C$ interacts with the nuclear pore complex and passes through the channel into the nucleus (compartment $B$).  

\item (Reaction 7, 4, 3) In the nucleus, RanGTP competes for binding with the receptor and causes the receptor to dissociate from the cargo. The new complex formed by RanGTP and receptor then translocates to the cytoplasm while the cargo is left inside the nucleus. 

\item (Reaction 2) Once in the cytoplasm, the GTPase activating protein (RanGAP) then binds to RanGTP, causing the hydrolysis of GTP to GDP and release of energy. 

\item (Reaction 1, 6) The RanGDP produced in this process then binds the nuclear transport factor NTF2 which returns it to the nucleus. 

\item (Reaction 5) Now in the nucleus, RanGDP interacts with a guanine nucleotide exchange factor (GEF) which replaces GDP with GTP, resulting again a RanGTP from, and beginning a new cycle.
\end{itemize}

\subsection{Kinetics equations}

The whole process can be formulated by a set of kinetics equations involving both cargo protein translocation and Ran regulation.  The molecular species in the kinetics equations are labelled according to Figure \ref{fig:model}.  
\\

\be
[X_1]+[X_2]\underset{k_1^-}{\stackrel{k_1^+}{\rightleftharpoons}}[X_3]
\ee
\be
[X_{6}]\underset{k_2^-}{\stackrel{k_2^+}{\rightleftharpoons}}[X_{7}]+[X_2]
\ee
\be
[X_{13}]\underset{k_3^-}{\stackrel{k_3^+}{\rightleftharpoons}}[X_{6}]
\ee
\be
[X_{14}]+[X_9]\underset{k_4^-}{\stackrel{k_4^+}{\rightleftharpoons}}[X_{13}]
\ee
\be
[X_{10}]\underset{k_5^-}{\stackrel{k_5^+}{\rightleftharpoons}}[X_{8}]+[X_9]
\ee
\be
[X_{3}]\underset{k_6^-}{\stackrel{k_6^+}{\rightleftharpoons}}[X_{10}]
\ee
\be
[X_{11}]\underset{k_7^-}{\stackrel{k_7^+}{\rightleftharpoons}}[X_{14}]+[X_{12}]
\ee
\be
[X_{7}]\underset{k_8^-}{\stackrel{k_8^+}{\rightleftharpoons}}[X_{14}]
\ee
\be
[X_{4}]\underset{k_9^-}{\stackrel{k_9^+}{\rightleftharpoons}}[X_{11}]
\ee
\be
[X_{7}]+[X_5]\underset{k_{10}^-}{\stackrel{k_{10}^+}{\rightleftharpoons}}[X_{4}]
\ee
\be
[X_{8}]\underset{k_{11}^-}{\stackrel{k_{11}^+}{\rightleftharpoons}}[X_{1}]
\ee

From this we can write down the following kinetics:
\bea
\frac{d [X_1]}{dt}&=&-k_1^+[X_1][X_2]+k_1^-[X_3]+k_{11}^+[X_8]-k_{11}^-[X_1]\\
\frac{d [X_2]}{dt}&=&-k_1^+[X_1][X_2]+k_1^-[X_3]+k_{2}^+[X_6]-k_{2}^-[X_7][X_2]\\
\frac{d [X_3]}{dt}&=&k_1^+[X_1][X_2]-k_1^-[X_3]-k_{6}^+[X_3]+k_{6}^-[X_{10}]\label{rateeqnX3}\\
\frac{d [X_4]}{dt}&=&-k_9^+[X_4]+k_9^-[X_{11}]+k_{10}^+[X_7][X_5]-k_{10}^-[X_4]\\
\frac{d [X_5]}{dt}&=&-k_{10}^+[X_5][X_7]+k_{10}^-[X_4]\\
\frac{d [X_6]}{dt}&=&-k_2^+[X_6]+k_2^-[X_2][X_7]+k_{3}^+[X_{13}]-k_{3}^-[X_6]\\
\frac{d [X_7]}{dt}&=&k_2^+[X_6]-k_2^-[X_2][X_7]-k_{8}^+[X_{7}]+k_{8}^-[X_{14}]-k_{10}^+[X_7][X_5]+k_{10}^-[X_4]\\
\frac{d [X_8]}{dt}&=&k_5^+[X_{10}]-k_5^-[X_8][X_9]-k_{11}^+[X_{8}]+k_{11}^-[X_1]\\
\frac{d [X_9]}{dt}&=&k_5^+[X_{10}]-k_5^-[X_8][X_9]-k_{4}^+[X_{14}][X_9]+k_{4}^-[X_{13}]\\
\frac{d [X_{10}]}{dt}&=&-k_5^+[X_{10}]+k_5^-[X_8][X_9]+k_{6}^+[X_{3}]-k_{6}^-[X_{10}]\label{rateeqnX10}\\
\frac{d [X_{11}]}{dt}&=&-k_7^+[X_{11}]+k_7^-[X_{12}][X_{14}]+k_{9}^+[X_{4}]-k_{9}^-[X_{11}]\\
\frac{d [X_{12}]}{dt}&=&k_7^+[X_{11}]-k_7^-[X_{12}][X_{14}]\\
\frac{d [X_{13}]}{dt}&=&-k_3^+[X_{13}]+k_3^-[X_6]+k_{4}^+[X_{14}][X_9]-k_{4}^-[X_{13}]\\
\frac{d [X_{14}]}{dt}&=&-k_4^+[X_{14}][X_9]+k_4^-[X_{13}]+k_{7}^+[X_{11}]-k_{7}^-[X_{14}][X_{12}]+k_8^+[X_7]-k_8^-[X_{14}]
\eea

\section{Estimating the rate constants}

Here we list the kinetics rate constants used in the simulation. Some of them are directly available from literature while others are estimated as described below. In the following, $a= 100\,\mu$m$^3$ s$^{-1}$ is the nuclear pore permeability and $\nu_N=100\, \mu$m$^3$ and $\nu_C=500\, \mu$m$^3$ are the nuclear and cytoplasm compartment volumes, respectively. The exponential free energy difference defined in Eq.\eqref{k5est}\eqref{k2est} are set to be: $e^{\Delta F}=e^{\Delta\tilde{F}}=50$. Note that volume factors modulate the permeabilities in the usual manner (see Eq.(1) in  \cite{michaelBiophyJ}): Namely, rate constants of cytosolic species (i.e. $X_1, X_3, X_4, X_6, X_7$) across the nuclear membrane is given by $k_\alpha^\pm = a/\nu_C$ with $\alpha = 3,6,8,9,11$ (i.e., reactions that involve crossing the nuclear pores). Rate constants for the nuclear counterparts (i.e. $X_{8}, X_{10}, X_{11}, X_{13}, X_{14}$) are, on the other hand, given by $k_\alpha^\pm = a/\nu_N$, with $\alpha = 3,6,8,9,11$. For example, kinetics equations for $X_3$ (cytosolic NTF2-RanGDP complex) and $X_{10}$ (nuclear NTF2-RanGDP complex) should read (c.f. Eq.\eqref{rateeqnX3} and Eq.\eqref{rateeqnX10}): 
\bea
\frac{d [X_3]}{dt}&=&k_1^+[X_1][X_2]-k_1^-[X_3]-\frac{a}{\nu_C}\left([X_3]-[X_{10}]\right)\\
\frac{d [X_{10}]}{dt}&=&-k_5^+[X_{10}]+k_5^-[X_8][X_9]+\frac{a}{\nu_N}\left([X_{3}]-[X_{10}]\right).
\eea

\begin{center}
\begin{tabular}[c]{|c|c|c|c|l|}\hline
reaction & $K_D$ or $k_{in}/k_{out}$& $k^+$ & $k^-$ & References and Note \\ \hline\hline
1 & 25 nM  & $0.1$ (nM$^{-1}$ s$^{-1}$)& 2.5 (s$^{-1}$)&  From  \cite{SmithScience}\\\hline
2 & $\sim$ & $e^{\Delta \tilde{F}}/ \theta$ (s$^{-1}$) & 1 (nM$^{-1}$ s$^{-1}$) &  estimate using Eq.\eqref{k2est}\\\hline
3 & $\sim$ & 1 or 0.2(s$^{-1}$) & 1 or 0.2(s$^{-1}$) & $k_3^\pm=a/\nu_N$ or $k_3^\pm=a/\nu_C$ \\\hline
4 & 10 nM & 0.1 (nM$^{-1}$ s$^{-1}$)& 1 (s$^{-1}$)& From  \cite{SmithScience}\\\hline
5 & $\sim$ & $e^{\Delta F}\times\theta$ (s$^{-1}$) & 1 (nM$^{-1}$ s$^{-1}$) & estimate using Eq.\eqref{k5est}\\\hline
6 & $\sim$ & 1 or 0.2   (s$^{-1}$) & 1 or 0.2 (s$^{-1}$) & $k_6^\pm=a/\nu_N$ or $k_6^\pm=a/\nu_C$  \\\hline
7 & 20 nM & $20$ (s$^{-1}$) & $1$ (nM$^{-1}$ s$^{-1}$) & From  \cite{HarremanJBC} \\\hline
8 & $\sim$ & 1 or 0.2  (s$^{-1}$)& 1 or 0.2(s$^{-1}$)& $k_8^\pm=a/\nu_N$ or $k_8^\pm=a/\nu_C$  \\\hline
9 & $\sim$ & 1 or 0.2  (s$^{-1}$)& 1 or 0.2(s$^{-1}$)& $k_9^\pm=a/\nu_N$ or $k_9^\pm=a/\nu_C$  \\\hline
10 & 20 nM & $1$ (nM$^{-1}$ s$^{-1}$) & $20$ (s$^{-1}$) & From  \cite{HarremanJBC} \\\hline
11 & $\sim$ & 1 or 0.2(s$^{-1}$)&  1 or 0.2(s$^{-1}$)& $k_{11}^\pm=a/\nu_N$ or $k_{11}^\pm=a/\nu_C$  \\\hline
\end{tabular}
\end{center}

\begin{center}
\begin{tabular}[c]{|c|c|}\hline
Labels & Species\\ \hline\hline
N  & NTF2\\ \hline
Im & Importin (importin)\\ \hline
RD & RanGDP \\ \hline
RT & RanGTP \\ \hline
N$\cdot$RD & NTF2+RanGDP complex \\ \hline
N$\cdot$RT & NTF2+RanGTP complex\\ \hline
Im$\cdot$RD & Importin+RanGDP complex \\ \hline
Im$\cdot$RT & Importin+RanGTP complex\\ \hline
fD  & (free) GDP \\ \hline
fT  & (free) GTP \\ \hline
\end{tabular}
\end{center}

\subsection{Reaction 5: Ran exchange mediated by RanGEF}

\begin{figure*}[!h]
\begin{center}
\includegraphics[width=8cm]{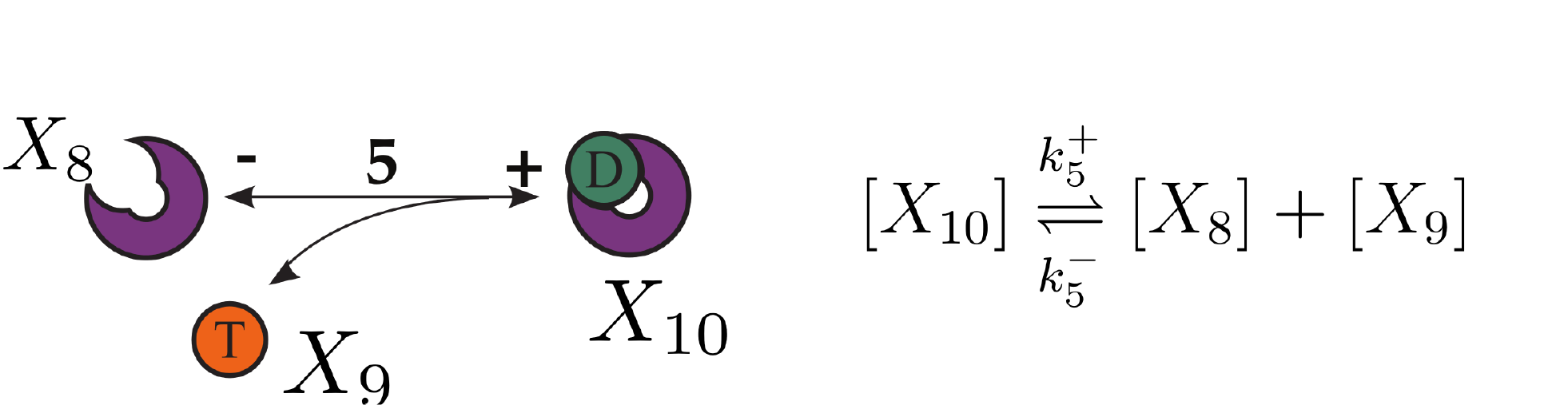}
\caption{{\bf Illustration of Ran GDP to GTP exchange reaction mediated by RanGEF }}
\label{reac5}
\end{center}
\end{figure*}

The goal is to estimate the $K_D$ for the following reaction:
\be
[N\cdot RD]\underset{k_5^-}{\stackrel{k_5^+}{\rightleftharpoons}}[N]+[RT],
\ee
namely, 
\be
\frac{k_5^+}{k_5^-}=\frac{[N][RT]}{[N\cdot RD]}
\ee
Consider the following two constituting reactions
\be\label{r51}
[N\cdot RD]+[fT]\underset{k_\alpha^-}{\stackrel{k_\alpha^+}{\rightleftharpoons}}[N\cdot RT]+[fD]
\ee

\be\label{r52}
[N\cdot RT]\underset{k_\beta^-}{\stackrel{k_\beta^+}{\rightleftharpoons}}[N]+[RT]
\ee
This implies (neglecting labels of steady states SS),
\bea\label{r51kd}
\frac{k_\alpha^+}{k_\alpha^-}&=&\frac{[N\cdot RT][fD]}{[N\cdot RD][fT]}\\ \label{r52kd}
\frac{k_\beta^+}{k_\beta^-}&=&\frac{[N][RT]}{[N\cdot RT]}
\eea
Thus we can reexpress Eq.\eqref{r51kd} using Eq.\eqref{r52kd}:
\be
\frac{k_\alpha^+}{k_\alpha^-}=\frac{1}{[N \cdot RD]}\frac{[fD]}{[fT]}\cdot\left(\frac{k_\beta^-}{k_\beta^+} [N][RT]\right)=\left(\frac{[N][RT]}{[N \cdot RD]}\right)\cdot \frac{[fD]}{[fT]}\frac{k_\beta^-}{k_\beta^+}=\frac{k_5^+}{k_5^-}\cdot\frac{[fD]}{[fT]}\frac{k_\beta^-}{k_\beta^+}
\ee

Thus
\be\label{k5est}
\frac{k_5^+}{k_5^-}=\frac{k_\alpha^+}{k_\alpha^-}\cdot\frac{k_\beta^+}{k_\beta^-}\cdot\frac{[fT]}{[fD]}\sim O(1)\cdot k_0 e^{\Delta F}\cdot \exp\left(\log\frac{[fT]}{[fD]}\right)
\ee
The first term (i.e. $k_\alpha^+/k_\alpha^-$ ) comes from guanine nucleotide exchange reaction and is of order one while the second (i.e. $k_\beta^+/k_\beta^-$) is related to the free energy difference between binding and un-binding of NTF2+RanGTP complex which is much larger than 1: $\Delta F>>1$. This can also be understood using Eq.\eqref{r52kd} by noting that in the nucleus NTF2 seldom binds to RanGTP. Finally, since the free GTP to GDP ratio, $[fT]/[fD]$, is buffered by cellular metabolism, we simply treat the last term as a free parameter $\theta$. Note that there is far more free GTP than Ran on a molar basis. After rescaling time by $\tau\leftarrow tc_0 k_{\text{diff}}$, with $k_\text{diff}=10$ sec$^{-1}$ nM$^{-1}$ and $c_0$ represent the diffusion-limited reaction rate and the characteristic molar concentration (set to 1nM), respectively, and approximating $e^{\Delta F}\approx 10\sim 100$ , one can estimate $(k_5^+/k_5^-)\sim (10\sim 100)\times \theta$, where $\theta:=[fT]/[fD]$ is treated as a free parameter. 

\subsection{Reaction 2: Ran exchange mediated by RanGAP} 

\begin{figure*}[!h]
\begin{center}
\includegraphics[width=8cm]{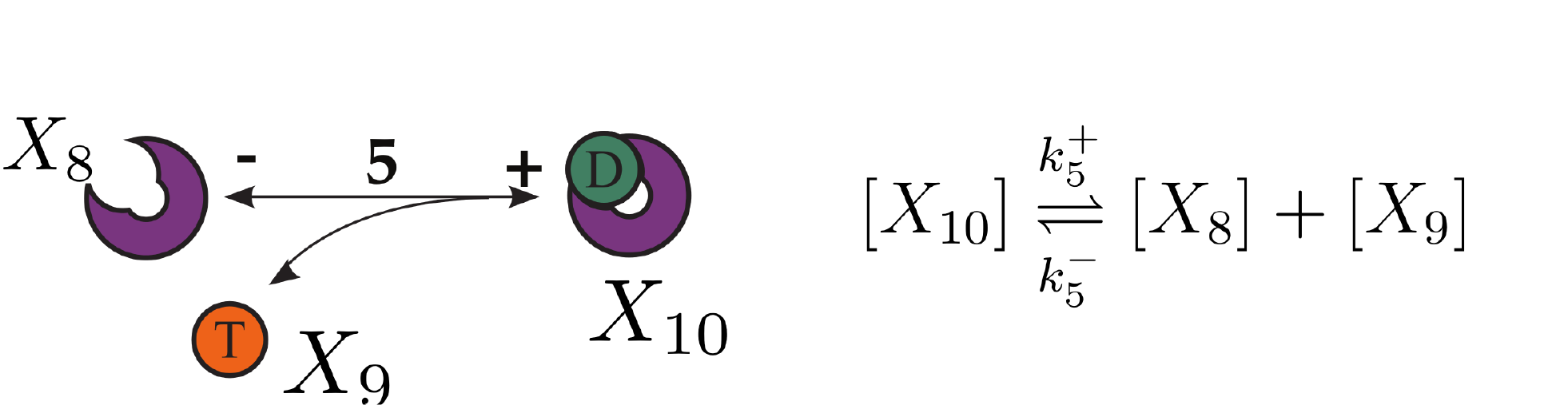}
\caption{{\bf Illustration of RanGTP to RanGDP exchange reaction mediated by RanGAP }}
\label{reac2}
\end{center}
\end{figure*}

We aim to approximate $K_D$ for such reaction:
\be
[Im\cdot RT]\underset{k_2^-}{\stackrel{k_2^+}{\rightleftharpoons}}[Im]+[RD],
\ee
\be
\frac{k_2^+}{k_2^-}=\frac{[Im][RD]}{[Im \cdot RT]}
\ee
Similarly the estimation is based on the following two steps:
\be\label{r21}
[Im \cdot RT]+[fD]\underset{k_\gamma^-}{\stackrel{k_\gamma^+}{\rightleftharpoons}}[Im \cdot RD]+[fT]
\ee

\be\label{r22}
[Im \cdot RD]\underset{k_\delta^-}{\stackrel{k_\delta^+}{\rightleftharpoons}}[Im]+[RD]
\ee
This implies (neglecting labels of steady states SS),
\bea\label{r21kd}
\frac{k_\gamma^+}{k_\gamma^-}&=&\frac{[Im\cdot RD][fT]}{[Im\cdot RT][fD]}\\\label{r22kd}
\frac{k_\delta^+}{k_\delta^-}&=&\frac{[Im][RD]}{[Im\cdot RD]}
\eea
Thus we can reexpress Eq.\eqref{r21kd} using Eq.\eqref{r22kd}:
\be
\frac{k_\gamma^+}{k_\gamma^-}=\frac{1}{[Im\cdot RT]}\frac{[fT]}{[fD]}\cdot\left(\frac{k_\delta^-}{k_\delta^+} [Im][RD]\right)=\left(\frac{[Im][RD]}{[Im\cdot RT]}\right)\cdot \frac{[fT]}{[fD]}\frac{k_\delta^-}{k_\delta^+}=\frac{k_2^+}{k_2^-}\cdot\frac{[fT]}{[fD]}\frac{k_\delta^-}{k_\delta^+}
\ee

Thus
\be\label{k2est}
\frac{k_2^+}{k_2^-}=\frac{k_\gamma^+}{k_\gamma^-}\cdot\frac{k_\delta^+}{k_\delta^-}\cdot\frac{[fD]}{[fT]}\sim O(1)\cdot k_0 e^{\Delta \tilde{F}}\cdot \exp\left(\log\frac{[fD]}{[fT]}\right)\sim k_0\times (10\sim 100)\times \frac{1}{\theta}
\ee


\section{Standard estimate of diffusion-limited reaction rate}

Considering two type of molecules A and B diffusing in a viscous environment. According the Fick's law the diffusion flux of one type of molecule assuming the other is at stationary is given as 
\be
\vec{J}_{\mu}=-D_\mu\nabla [\mu],
\ee
where $\mu=A,B$ and $D_\mu$ is the diffusion constant of molecule $\mu$. Assuming spherical symmetry one can integrate Fick's law to get the total number of molecules diffusing through a given surface area:
\be
\phi_{tot}=4\pi R (D_A+D_B)[A][B],
\ee
where $R$ is the sum of molecular radii of A and B. The factor $k_a:=4\pi R (D_A+D_B)$ is exactly the reaction rate of the overall catalytic reaction under the assumption that the process is diffusion-limited (i.e. upon A and B are in contact, the intermediate complex AB immediately reacts to form the final product P):
\be
A+B\xrightarrow{k_a}P
\ee
Finally, recall Stokes-Einstein relation: $D=k_BT/(6\pi\eta a)$ with molecule (spherical) particle radius $a$, we have
\be
k_a=4\pi (2a) \left(2\times \frac{k_BT}{6\pi\eta a}\right) \left[\frac{\text{m}^3}{\text{sec}}\right]\rightarrow \left(\frac{8k_BT}{3\eta }\right)N_A 10^3  \left[\frac{1}{\text{M}\cdot\text{sec}}\right],
\ee
where $N_A$ is the  Avogadro's constant. The factor $10^3$ appears because we convert the SI unit of volume m$^3$ to liter.  
Using $\eta=10^{-1}$ (Pa$\cdot$sec)= $10^{-3}$ kg/m/sec, we get
\be
k_{\text{diff}}:=k_a\sim 10\times 10^9 \,[\text{M}^{-1}\cdot\text{ sec}^{-1}]= 10 \,[\text{nM}^{-1}\cdot\text{ sec}^{-1}]
\ee

\section{Simulation codes}

MATLAB$^\circledR$ simulation codes are available for download at \url{http://physics.bu.edu/~chinghao/thermo_transport/codes/}

\section{Entropy Production}

The distinct feature of systems out of thermodynamics equilibrium is the continuous production of entropy. The rate of entropy change (in time) consists of two parts: (i) the internal entropy change and (ii) the exchange of entropy with the environment 
\be
\frac{dS}{dt}=\Pi-\Phi,
\ee
where $S$ is the entropy of the system and $\Pi$ is the rate of entropy production and $\Phi$ denotes the rate of entropy flow from the system to the outside. Within this context, the 2nd law of thermodynamics dictates $\Pi\ge 0$ and the notion of steady states translates into $\Pi=\Phi$: entropy produced is continuously given away to the environment. One can further distinguishes the equilibrium from the nonequilibrium steady states by
\begin{itemize}
\item Equilibrium steady states (ESS): $\Pi=\Phi=0$
\item Nonequilibrium steady states (NESS, i.e. irreversible): $\Pi=\Phi>0$
\end{itemize}

Consider systems that can be described by a continuous time Markov process such that the probability flow can be written as a master equation:
\be\label{master}
\frac{d}{dt} P_i(t)=\sum_j [P_j(t)W_{ji}-P_i(t)W_{ij}],
\ee
where $W_{ij}$ is the transition rate from state $j$ to state $i$ and $P_i(t)$ is the probability of state $i$ at time $t$. An appropriate microscopic description for the nonequilibrium system  amounts to (i) having well-defined entropy for the irreversible systems and (ii) the entropy production rate $\Pi$ should respects the non-negativity and should vanish when system equilibrates (i.e. when it exhibits reversibility). For systems described by the master equation, thermodynamics equilibrium is essentially the detailed-balanced condition: $P_iW_{ij}=P_jW_{ji}$. The solution for the first is the Boltzmann-Gibbs entropy:
\be\label{ent}
S(t)=-k_B\sum_{i}P_i(t)\log P_i(t),
\ee
while the entropy production rate is advanced by the Schnakenberg description \cite{NESSref}:
\be\label{nonequpi}
\Pi(t)=\frac{k_B}{2}\sum_{ij}[P_i(t)W_{ij}-P_j(t)W_{ji}]\log\frac{W_{ij}P_j(t)}{W_{ji}P_i(t)}
\ee

By imposing $d S(t)/dt =0$ to Eq.\eqref{ent} at steady state and using Eq.\eqref{master}\eqref{nonequpi} to simply, one gets the steady state entropy production rate:
\be\label{equpi}
\Pi=k_B\sum_{ij} W_{ij}P_j\log\frac{W_{ij}}{W_{ji}},
\ee
where $P_i$ is the stationary probability distribution. It's easy to check that $\Pi-\Pi(t)=dS/dt\rightarrow 0$ in the stationary state.

One can map the network in FIG.~\ref{fig:model}B to a nonequilibrium Markov process. A non-equilibrium steady state (NESS) essentially necessitates breaking the detailed balance in the underlying Markov process and therefore, the system has a nonzero entropy production that is continuously given away to the environment. Such entropy production is exactly the power consumed by the circuit to maintain NESS. Now defining $EP:=\Pi\times T $ using \eqref{equpi} (in the same spirit as $F=U-TS$, where $F$ is the Helmholtz free energy), we have
\be\label{EPSI}
EP=k_B T\sum_{i,j}P_i^{SS}W(i,j)\log\frac{W(i,j)}{W(j,i)},
\ee 
where $P_i^{SS}$ is the steady state probability distribution of state $i$ while $W(i,j)$ denotes the transition probability from state $i$ to state $j$. Concretely, $P_i^{SS}$ is the fraction of reactants participating in the transition reaction starting from state $i$ while $W(i,j)$ can be calculated from the relevant reaction fluxes. For example, $P_3^{SS}$ is the molar fraction of cytoplasmic NTF2-RanGDP ($\sim [X_3]$) whereas $W(3,10)$ is the transition probability of of NTF2-RanGDP into the nucleus: $W(3,10)=(k_6^+ [X_3])/(k_6^+[X_3]+k_1^-[X_3])$ (See FIG.\ref{fig:model}C). Note that in principle the summation in Eq.\eqref{EPSI} to obtain the entropy production is taken over all links in FIG.\ref{fig:model}B. It can be separated, however, into reactions 1-6 that represent the Ran futile cycle (i.e. FIG. \ref{fig:model}C) and the remaining reactions 7-11 that do not explicitly involve Ran. The latter are essentially passive and could be expected to satisfy detailed balance at steady state. We have confirmed numerically that the contributions of reactions 7-11 in Eq.\eqref{EPSI} cancel to zero, so the total entropy production is equal that evaluated in the futile cycle alone. We can also inspect the reactions qualitatively. Reaction 9 is trivially in detailed balance because the concentrations $X_4$ and $X_{11}$ are equal in steady state. These represent the importin-cargo complex in cytoplasm and nucleus, respectively. Clearly the net cargo binding/unbinding to importin in the cytoplasm must balance that in the nucleus as well, so the contributions of reactions 7 and 10 cancel. Finally, the free receptors importin and NTF2 exchange passively across the nuclear envelope (reactions 8 and 11). Again in steady state their cycle fluxes must balance, so their contributions to the entropy production sum also cancel.

\section{Weak sensitivity to the GTP:GDP ratio $\theta$}

Ultimately the (chemical) free-energetic fuel driving the transport cycle is the ratio of GTP to GDP, $\theta$, which is held out of equilibrium by cellular metabolism. We find that the nuclear localization ratio, as well as biased concentrations of transport receptors, is not strongly dependent on $\theta$. This reflects the counterbalancing effects of RanGEF and RanGAP as described in Section II above.

\begin{figure*}[!h]
\begin{center}
\includegraphics[width=16cm]{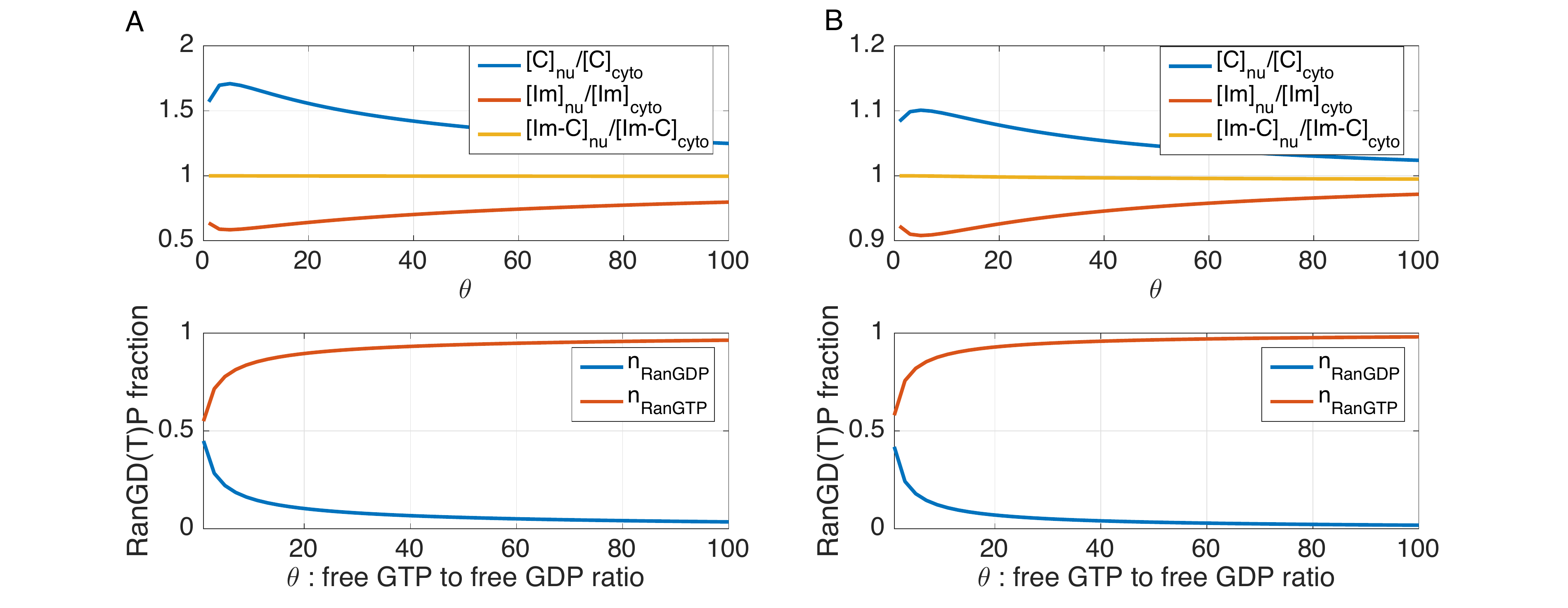}
\caption{{\bf Effect of free GTP to free GDP ratio $\theta$ on nuclear localization}. (A) [NTF2]$_{tot}$=100 nM (B) [NTF2]$_{tot}$=10 nM. Other parameters are the same for both panels:  [C]$_{tot}$=10 nM, [Ran]$_{tot}$=75 nM and [Im]$_{tot}$=100 nM. Kinetics rate constants used are given in SM Section II.}
\label{fig:thetaeffect}
\end{center}
\end{figure*}

\begin{figure*}[!h]
\begin{center}
\includegraphics[width=16cm]{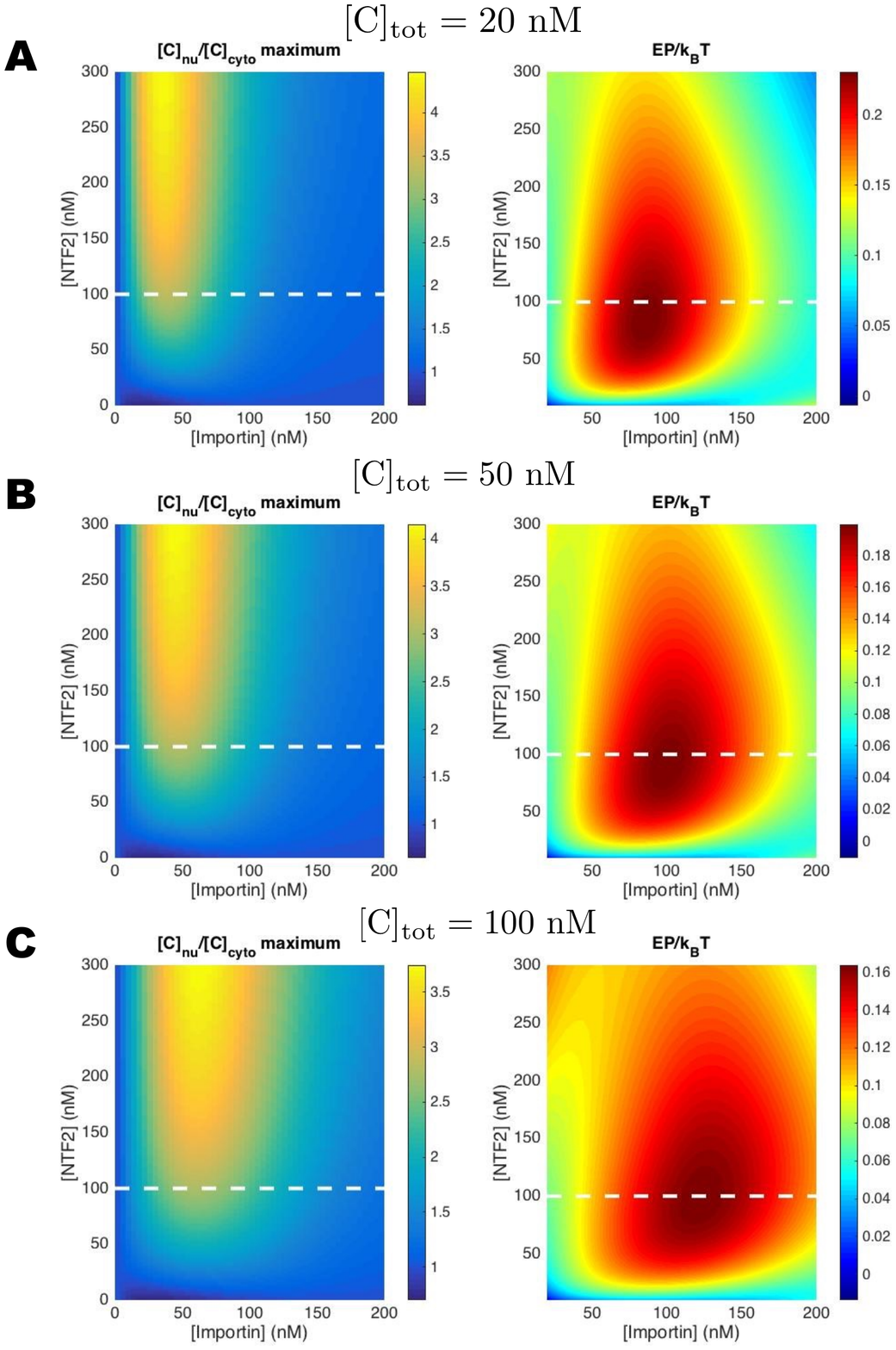}
\caption{{\bf Phase diagram of nuclear localization and entropy production}. (A) total cargo concentration [C]$_{\text{tot}}=20$ nM (B) [C]$_{\text{tot}}=50$ nM and (C) [C]$_{\text{tot}}=100$ nM. Other parameters used are the same as in FIG.3 and 4: [Ran]$_{\text{tot}}=75$ nM.  }
\label{fig:phaseAllC}
\end{center}
\end{figure*}

%
\end{document}